\documentclass[aps,preprint,showpacs,floatfix]{revtex4-2}

\usepackage{graphicx}
\usepackage{amssymb}
\usepackage{amsmath}
\usepackage{subfig}

\def\bea{\begin{eqnarray}}
\def\eea{\end{eqnarray}}
\def\beq{\begin{equation}}
\def\eeq{\end{equation}}

\def\bm{\begin{math}}
\def\me{\end{math}}

\begin{document}
\title{Aging Dynamics and Velocity Field Correlations in Three-Dimensional Uniformly Heated Granular Gases: A Molecular Dynamics Study}
\author{Rameez Farooq Shah$^{1}$, Shikha Kumari$^{2}$, Syed Rashid Ahmad$^{1}$} 
\email{rmzshah@gmail.com,\\
shikha1011@gmail.com,\\
srahmad@jmi.ac.in}
\affiliation{
$^{1}$Department of Physics, Jamia Millia Islamia, New Delhi 110025, India \\
$^{2}$Department of Physics, School of Basic and Applied Sciences,
IILM university, Greater Noida, Uttar Pradesh 201306
}

\begin{abstract}
We conduct a molecular dynamics simulation of an inelastic gas system utilizing an event-driven algorithm combined with a thermostat mechanism. Initially, the kinetic energy of the system experiences a decay before settling into a non-equilibrium steady state. To explore the aging characteristics, we analyze the velocity autocorrelation function, denoted as \( C(t_w, t) \).
Our findings indicate that \( C(t_w, t) \) exhibits a dependence on both waiting time \( t_w \) and correlation time \( t \) in an independent manner. At the outset, \( C(t_w, t) \) demonstrates an exponential decay pattern. With increasing \( t_w \), a slower decay is observed, which can be attributed to the development of correlations in the velocity field.
The explicit relationship of \( C(t_w, t) \) with respect to \( t_w \) serves as compelling evidence of the aging properties present in the system. These results deepen our comprehension of non-equilibrium statistical mechanics and the dynamics of dissipative systems.
Our research has significant implications for a range of applications involving inelastic collisions, extending from granular materials to phenomena observed in astrophysics. The simulation methodology and the insights gained contribute to the wider field of complex systems, shedding light on the behavior of systems that are far from equilibrium, particularly those characterized by energy dissipation due to inelastic interactions.
\end{abstract}
\maketitle
\section{\label{introd} INTRODUCTION}
Granular materials—composed of macroscopic particles that interact through dissipative collisions—has gained considerable interest in recent decades, owing to their wide-ranging applications across various industries and natural processes. These materials exhibit unique characteristics that resemble both fluids and solids \cite{rmp_behringer, rmp_kadanoff}. In their solid state, they can form heaps and resist deformation, exemplified by a pile of sand that remains stable at rest. Conversely, dry sand or powders can flow through an hourglass neck similar to a liquid. Furthermore, when subjected to external agitation, dry sand can behave like a gas. 
Granular materials typically consist of polydisperse particles in terms of size and shape, generally exceeding $1 \mu$m and existing in arbitrary forms. Their macroscopic dimensions imply that they are unaffected by thermal fluctuations. In theoretical and numerical analyses, these particles are frequently represented as spheres, needles, or cylinders \cite{rmp_tsimring, duran, ristow, nb_ktgg}. A notable characteristic of granular materials is the dissipative nature of the interactions between particles, resulting in a loss of kinetic energy or cooling, which is often accompanied by a local alignment of particle velocities. This dissipation leads to various fascinating phenomena, such as size segregation, clustering, pattern formation, inelastic collapse, and anomalous velocity distributions \cite{haff83, swinney9596, gz93, mcny9296, jjbrey9698, tpcvn9798, sl9899, ap0607, adsp1213}.
Granular gas, also known as dilute granular systems, serves as a framework for exploring the characteristics of a gas where molecular interactions lead to energy dissipation. The investigation of a granular gas starts with the evolution of uniformly distributed inelastic particles. In the absence of external energy input, the system experiences a loss of kinetic energy due to inelastic collisions among the particles. Initially, the density appears uniform, and the system enters a homogeneous cooling state (HCS). However, due to fluctuations in density and velocity fields, the HCS becomes unstable, transitioning the system into an inhomogeneous cooling state (ICS) \cite{haff83, ap0607, dp03}. In the ICS, regions with dense particle clusters emerge and continually expand, with particles in a cluster moving in nearly parallel directions. In experimental contexts, energy loss is often balanced by external energy input via various driving mechanisms, such as horizontal or vertical vibrations or rotation. Under these conditions, the system attains a nonequilibrium steady state \cite{swinney9596, ristow}.
A crucial element in the study of granular gases is the analysis of velocity autocorrelations and their aging characteristics. In this paper, we investigate the aging phenomena associated with the velocity autocorrelations of a hard sphere granular gas that has been uniformly heated using a Gaussian white noise thermostat. The investigation of heated granular gases has been extensively addressed through both analytical and computational approaches \cite{vne98}. To maintain energy in these systems, several thermostat mechanisms have been utilized. Our research employs the algorithm developed by Williams et al., which incorporates a white-noise thermostat (WNT) to uniformly heat the particles \cite{willmac96, william96}. This technique enables us to track the evolution of velocity autocorrelations over time and examine the aging effects typical of these non-equilibrium systems. By concentrating on the aging behavior of velocity autocorrelations, we aim to enhance the understanding of the complex dynamics present in uniformly heated granular gases and contribute to the wider field of non-equilibrium statistical mechanics.
The structure of the paper is as follows. In Section \ref{themodel}, we provide a thorough description of our model for a uniformly heated hard sphere granular gas, emphasizing the mechanisms that facilitate the exploration of aging phenomena in velocity autocorrelations. We detail the application of the white-noise thermostat to maintain a constant energy input and discuss how this configuration enables us to observe and analyze the non-equilibrium behavior of the system over prolonged time scales. In Section III, we present detailed results from our molecular dynamics simulations, focusing on the velocity autocorrelations of the heated hard sphere granular gas. Finally, Section IV summarizes our findings and discusses their implications for understanding the velocity statistics in driven granular systems.
%\newpage
\section{\label{themodel} Formalism}
We begin with a uniform granular gas composed of identical spherical particles. For simplicity, we assume the mass and diameter of these particles to be one. For hard spheres, the velocities of particles labeled \( i \) and \( j \) after a collision can be expressed in terms of their pre-collision velocities using the following relations:
\begin{align}
\label{coll_modified}
\vec{v}_i' &= \vec{v}_i - \frac{1 + r}{2} \left[\hat{n} \cdot (\vec{v}_i - \vec{v}_j)\right] \hat{n}, \\
\vec{v}_j' &= \vec{v}_j + \frac{1 + r}{2} \left[\hat{n} \cdot (\vec{v}_i - \vec{v}_j)\right] \hat{n},
\end{align}
where \( r (< 1) \) denotes the coefficient of restitution, and \( \hat{n} \) is a unit vector directed from the position of particle \( j \) to that of particle \( i \).
Similar to molecular gases, we can define a temperature for granular gases, termed granular temperature, represented as \( T = \frac{\langle \vec{v}^2 \rangle}{d} \), where \( \langle \vec{v}^2 \rangle \) signifies the mean-squared velocity and \( d \) indicates the dimensionality. In the initial stages, absent external driving forces, the rate of change of granular temperature can be described as follows \cite{haff83}:
\begin{equation}
\label{cool_modified}
\frac{dT}{dt} = -\frac{\epsilon \omega(T) T}{d}, \quad \epsilon = 1 - r^2,
\end{equation}
where \( \omega(T) \) denotes the collision frequency at temperature \( T \). From the kinetic theory of gases, we can express \( \omega(T) \) as \cite{cc70}:
\begin{equation}
\label{omega_modified}
\omega(T) \approx \frac{2 \pi^{(d-1)/2}}{\Gamma(d/2)} \chi(n) n T^{1/2},
\end{equation}
with \( \chi(n) \) being the pair correlation function at contact for hard spheres with density \( n \). Utilizing equations~(\ref{cool_modified}) and (\ref{omega_modified}), we derive Haff's law for the homogeneous cooling state:
\begin{figure}
    \centering
    \includegraphics[scale=0.55]{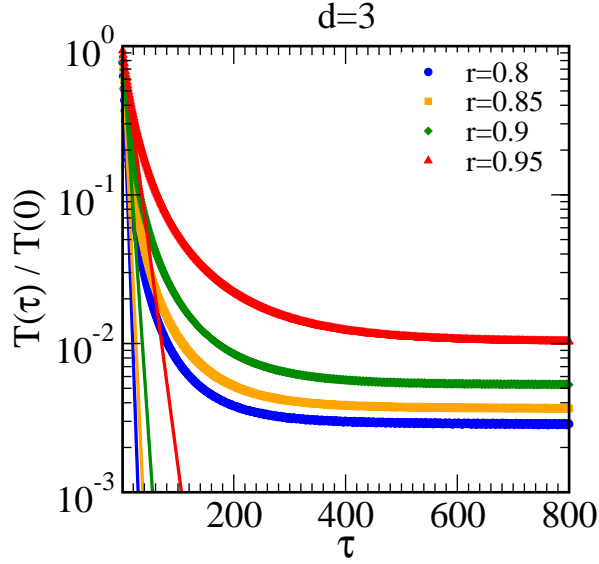} 
    \caption{Time dependence of the granular temperature in \( d = 3 \), presented on a semilog scale. The normalized granular temperature \( T(\tau)/T(0) \) is plotted against \( \tau \) for \( r = 0.80, 0.85, 0.90, \) and \( 0.95 \). The solid lines represent Haff's law.}
    \label{fig_haff1}
\end{figure}
\begin{equation}
T(t) = T_0 \left(1 + \frac{\epsilon \omega(T_0)}{2d} t\right)^{-2},
\end{equation}
where \( T_0 \) represents the initial temperature. 
Defining the average number of collisions that occur within a time interval \( t \) as \( \tau \), we have:
\begin{align}
\tau(t) &= \int_0^t dt' \, \omega(t') \nonumber \\
&= \frac{2d}{\epsilon} \ln\left(1 + \frac{\epsilon \omega(T_0)}{2d} t\right).
\end{align}
As energy dissipates from the system, the number of collisions increases logarithmically over time, rather than linearly. Expressing Haff's law in terms of \( \tau \), we obtain:
\begin{equation}
\label{haff_modified}
T(\tau) = T_0 \exp\left(-\frac{\epsilon}{d} \tau\right).
\end{equation}

\begin{figure}
    \centering
    \includegraphics[scale=0.55]{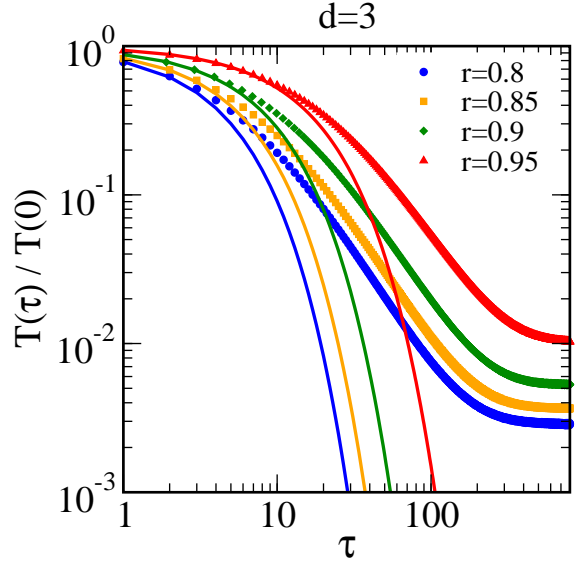} 
    \caption{Time dependence of the granular temperature in \( d = 3 \), displayed on a log-log scale. It is observed that the temperatures for each value of \( r \) stabilize to a constant value.}
    \label{fig_haff2}
\end{figure}
When external forces are applied, the energy supplied by these forces counteracts the energy lost during particle collisions, allowing the system to reach a steady non-equilibrium state. For a driven granular system, the stochastic motion of particles is governed by the following equation of motion:
\begin{equation}
    m \frac{d \mathbf{v}_i}{dt} = \mathbf{F}_i^{c} + \mathbf{F}_i^{t}
\end{equation}
Here, \( m \) denotes the mass of the particle, \( \mathbf{F}_i^{c} \) represents the net force acting on the \( i^{th} \) particle (where \( i = 1, 2, \ldots, N \)) due to pairwise collisions as described in the previous relations, and \( \mathbf{F}_i^{t} \) signifies the external force modeled as Gaussian white noise with zero mean. The correlation properties of the stochastic force are expressed as:
\begin{equation}
    \langle F_{i,\alpha}^{t}(t) F_{j,\beta}^{t}(t') \rangle = \xi_0^2 \delta_{ij} \delta_{\alpha \beta} \delta(t - t')
\end{equation}

\begin{equation}
    \langle \mathbf{F}_i^{t}(t) \rangle = 0
\end{equation}
where \( \alpha, \beta \in \{x, y, z\} \), \( \xi_0 \) quantifies the intensity of the stochastic force, \( \delta_{ij} \) and \( \delta_{\alpha \beta} \) represent the Kronecker delta, and \( \delta(t - t') \) is the Dirac delta function.

\begin{figure}[h]
    \centering
    \begin{minipage}[b]{0.7\textwidth}
        \centering
        \includegraphics[width=\textwidth]{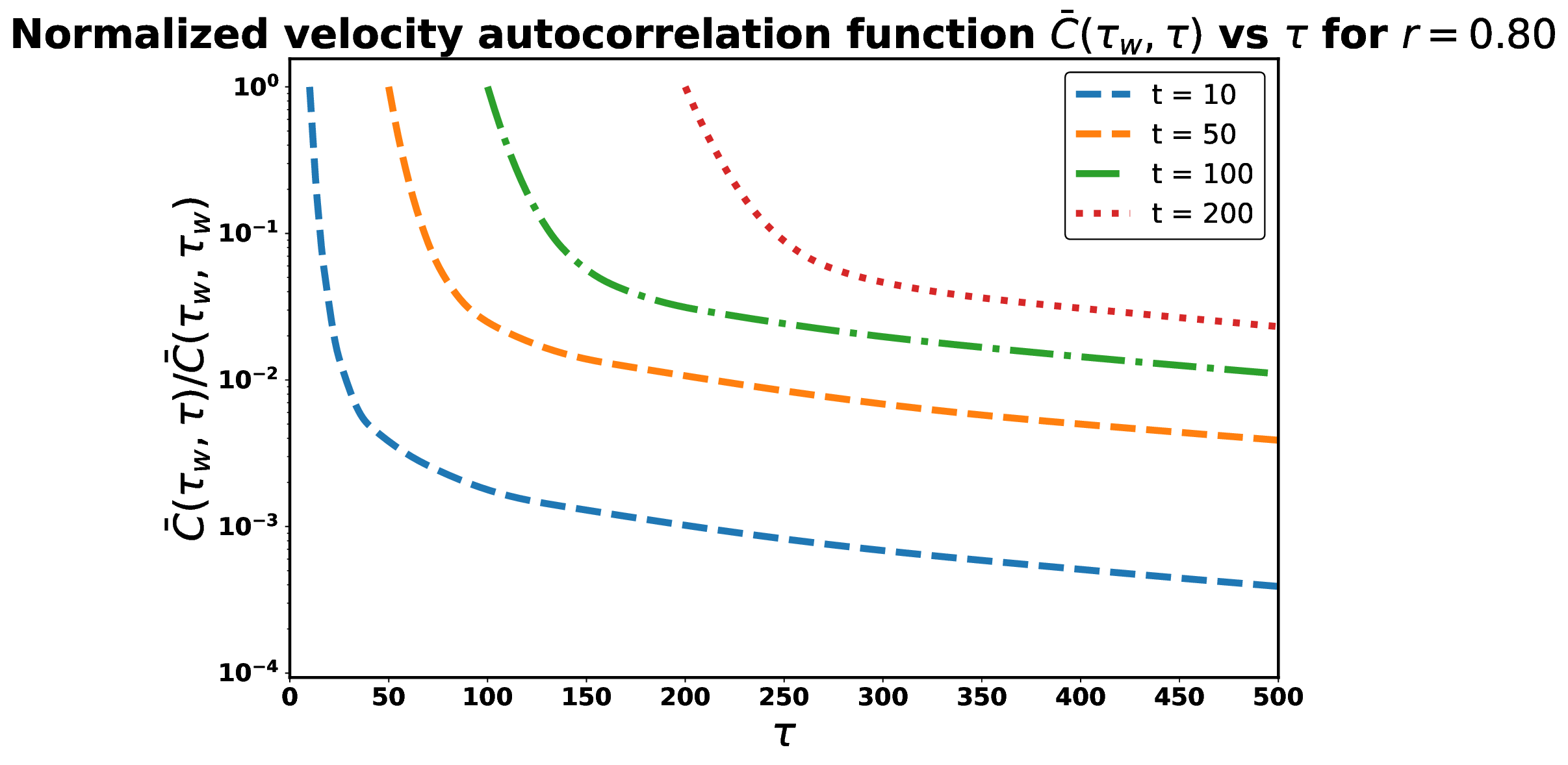}
        \caption*{(a) $r = 0.80$}
        \label{fig:subfig1}
    \end{minipage}
    \hspace{0.05\textwidth} % Adjust space between figures
    \begin{minipage}[b]{0.7\textwidth}
        \centering
        \includegraphics[width=\textwidth]{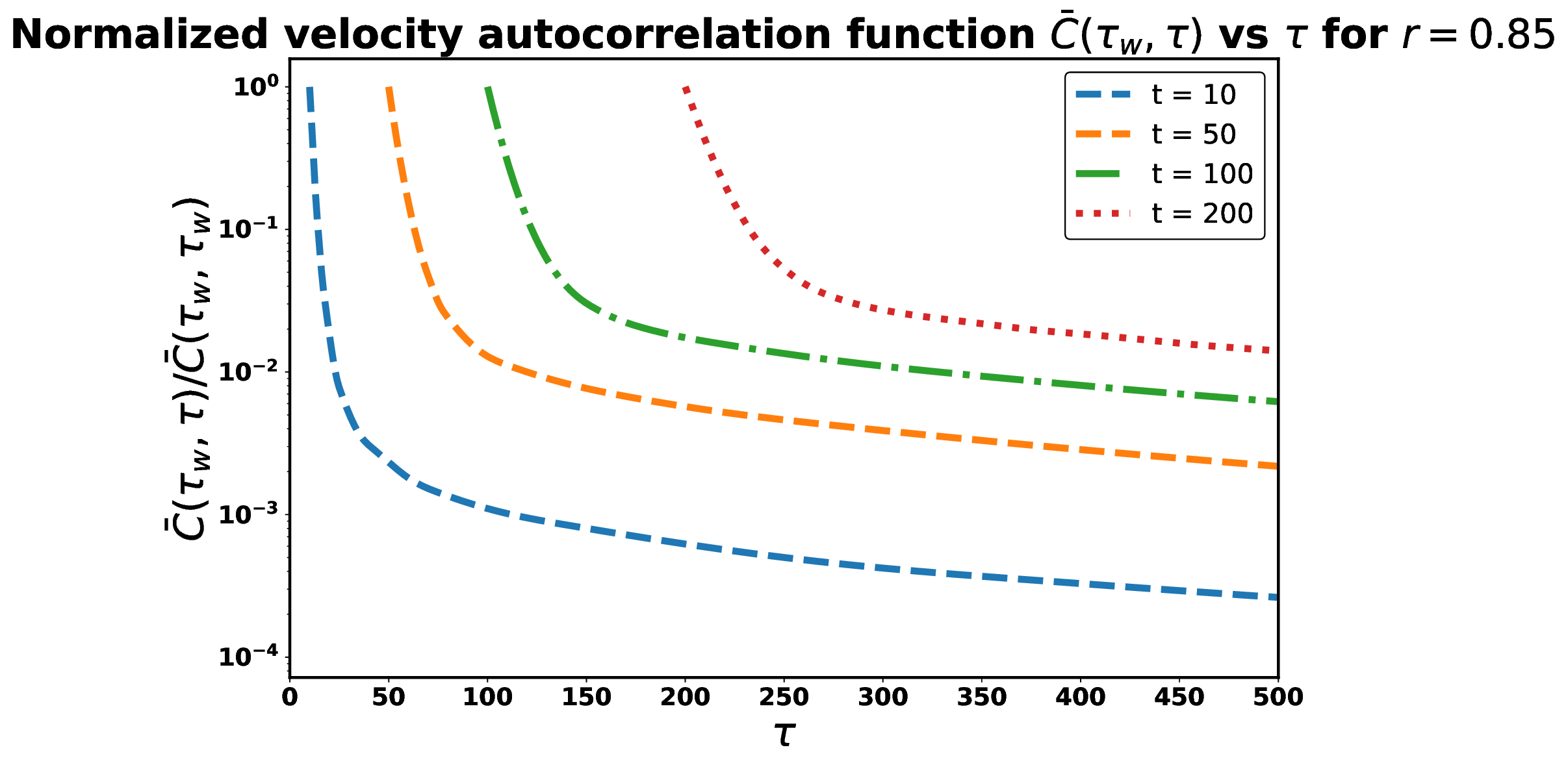}
        \caption*{(b) $r = 0.85$}
        \label{fig:subfig2}
    \end{minipage}
    \caption{The normalized velocity autocorrelation function \( \bar{C}(\tau_w, \tau) \) versus \( \tau \) for different values of \( r \). Plots \textbf{(a)} and \textbf{(b)} correspond to \( r = 0.80 \) and \( 0.85 \), respectively.}
    \label{fig_VDF_1}
\end{figure}
 
\section{\label{vdf} Velocity Autocorrelation Function}
The velocity autocorrelation function quantifies how the velocities of particles at different times are correlated. It is defined as follows:
\begin{equation}
C(\tau_w, \tau) = \frac{1}{N} \sum_{i=1}^{N} \mathbf{v}_i(\tau_w) \cdot \langle \mathbf{v}_i(\tau) \rangle
\end{equation}
Here, \( C(\tau_w, \tau) \) represents the velocity autocorrelation function computed at time \( \tau \) with respect to the waiting time \( \tau_w \). This function serves as a tool to investigate the temporal memory of the system’s behavior. During collisions, the velocities of neighboring particles become parallelized, leading to emergent correlations within the velocity field. This correlated motion facilitates the development of clusters, characterized by regions rich in particles as well as those deficient in particles. The stability and longevity of these clusters are captured by the velocity autocorrelation function.
In the context of inelastic collisions occurring in a homogeneous cooling state (HCS), the dependence of \( C(\tau_w, \tau) \) has been established in the work of Ben-Naim and Krapivsky (2002). Specifically, the autocorrelation function in HCS exhibits the following decay behavior:
\begin{equation}
C(\tau_w, \tau) = C(\tau_w, \tau_w) \exp \left( \frac{1 + \epsilon}{2d} (\tau - \tau_w) \right)
\end{equation}

\begin{figure}[h]
    \centering
    \begin{minipage}[b]{0.7\textwidth}
        \centering
        \includegraphics[width=\textwidth]{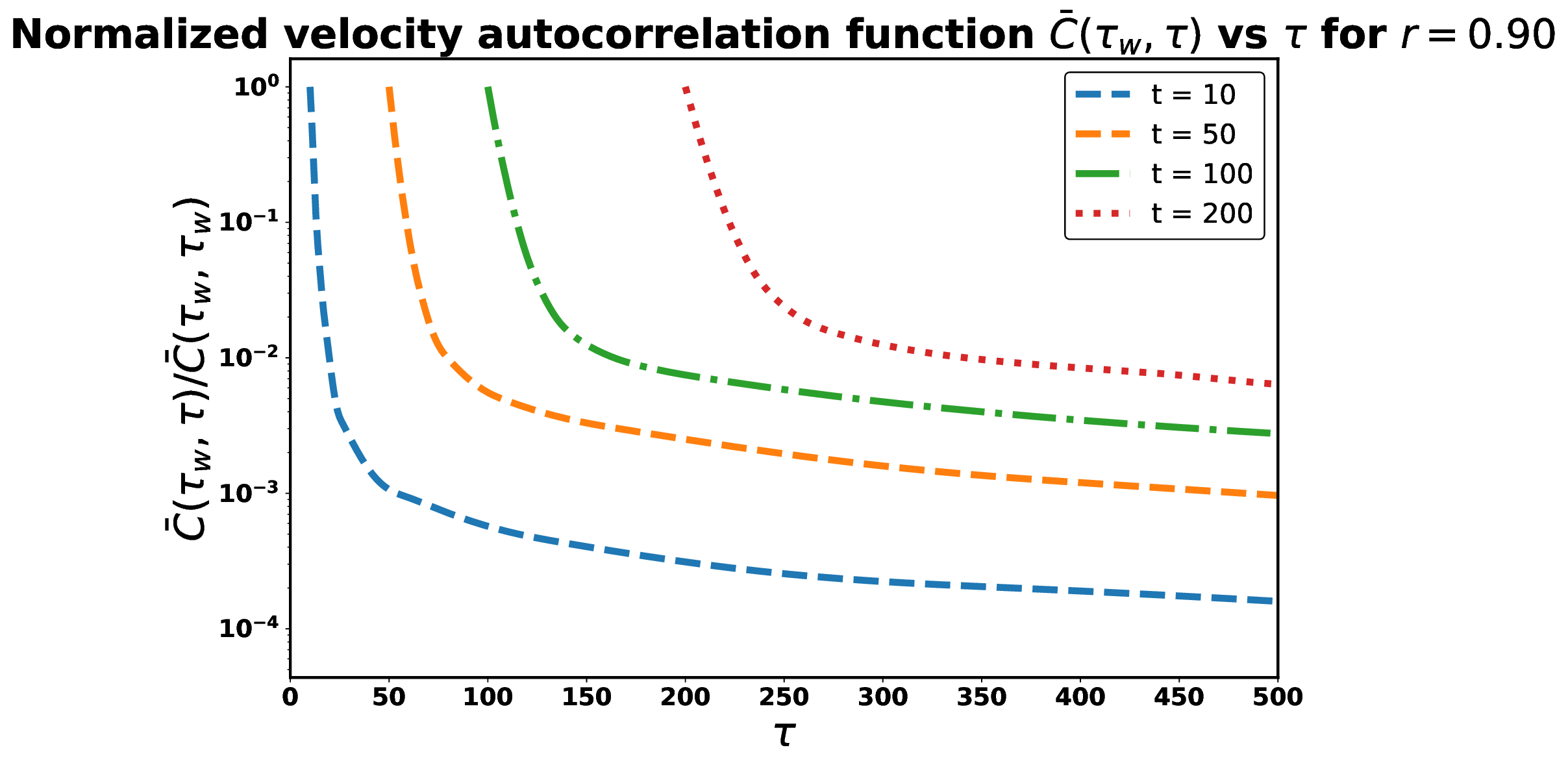}
        \caption*{(c) $r = 0.90$}
        \label{fig:subfig3}
    \end{minipage}
    \hspace{0.05\textwidth} % Adjust space between figures
    \begin{minipage}[b]{0.7\textwidth}
        \centering
        \includegraphics[width=\textwidth]{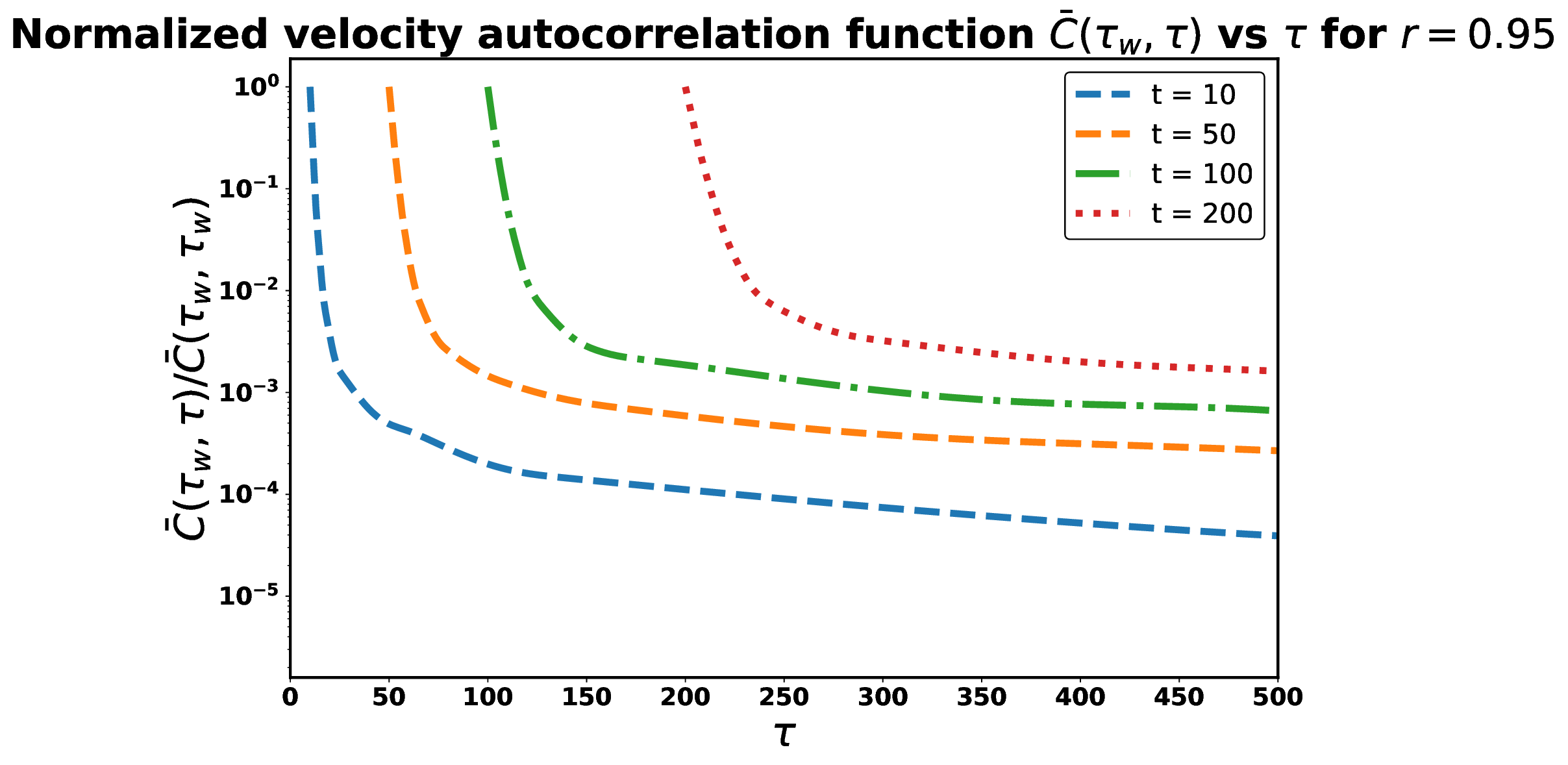}
        \caption*{(d) $r = 0.95$}
        \label{fig:subfig4}
    \end{minipage}
    \caption{The normalized velocity autocorrelation function \( \bar{C}(\tau_w, \tau) \) versus \( \tau \) for different values of \( r \). Plots \textbf{(c)} and \textbf{(d)} correspond to \( r = 0.90 \) and \( 0.95 \), respectively.}
    \label{fig_VDF_2}
\end{figure}
Alternatively, it can be expressed as:
\begin{equation}
C(\tau_w, \tau) = dT(0) \exp \left( \frac{1 - \epsilon^2}{2d(1 + \epsilon)} \tau_w \exp \left( - \frac{\tau - \tau_w}{d} \right) \right)
\end{equation}
For a system in equilibrium, the autocorrelation function \( C(\tau_w, \tau) \) depends solely on the time difference \( (\tau - \tau_w) \). In contrast, for a non-equilibrium system, the relationship between \( C(\tau_w, \tau) \) and the time difference is more complex; it does not follow a simple dependence on \( (\tau - \tau_w) \) but instead relies independently on both \( \tau_w \) and \( \tau \). The influence of \( \tau_w \) on \( C(\tau_w, \tau) \) is termed aging. Additionally, the autocorrelation function is affected by the dissipation present in the system. In the scenario of completely elastic collisions—where there is no energy dissipation—\( C(\tau_w, \tau) \) rapidly decays to zero, typically within just a few collisions per particle.
\section{\label{details_results} SIMULATION DETAILS AND RESULTS}
The system begins by assigning each particle a random position and velocity. It comprises \( N = 500,000 \) particles enclosed in a 3D cubic box with periodic boundary conditions, 

resulting in a number density of \( n = 0.02 \). The initial configuration ensures that the cores of no two particles overlap. The random velocity components are selected such that \( \Sigma \vec{v}_i = 0 \). The system evolves to \( \tau = 100 \) at \( r = 1 \) without additional energy input, guaranteeing that the system relaxes to a Maxwell-Boltzmann velocity distribution. This state serves as the initial condition for our simulation.
We then continue the evolution of the system until \( \tau = 1000 \) for four different values of the restitution coefficient \( r \) (\( r = 0.95, 0.90, 0.85, \) and \( 0.80 \)). The results presented correspond to averages over \( 50 \) independent initial conditions.
To simulate a system of inelastic hard sphere particles, we employed event-driven molecular dynamics \cite{allen, rapaport}. All particles are identical, with a unit mass \( m = 1 \) and diameter \( \sigma = 1 \). The post-collision velocities are derived from the pre-collision velocities using the relation shown in Equation 1. The system experiences Gaussian white noise, where the heat component is added to the velocity of each molecule at each time step \( dt \) according to the following equation:
\begin{equation}
    v_i(t+dt) = v_i(t) + \sqrt{A} \sqrt{dt} \xi. 
\end{equation}
Here, \( \xi \) is a random variable uniformly distributed in the interval \( \left[ -\frac{1}{2}, \frac{1}{2} \right] \), and \( A \) is the noise amplitude, set to \( 0.001 \). After the velocities are adjusted, the system is transformed to the center of mass frame to ensure the conservation of linear momentum:
\begin{equation}
    \textbf{v}_i = \textbf{v}_i - \frac{1}{N} \sum_{i=1}^{N} \textbf{v}_i. 
\end{equation}
At \( t = 0 \), the system is initialized with a homogeneous density field where the velocity components follow the Maxwell-Boltzmann distribution. Due to dissipative collisions, particles lose kinetic energy, and the thermostat noise partially compensates for this energy loss. This process continues until a state is reached where the losses from dissipation are counterbalanced by the gains from noise. In Fig. \ref{fig_haff1}, we present the time evolution of the reduced temperature \( T(\tau)/T(0) \) as a function of \( \tau \) on a semilog scale for various values of the restitution coefficient \( r \). Haff's law is also shown as solid lines for reference.
Fig. \ref{fig_haff2} illustrates the time evolution of the reduced temperature \( T(\tau)/T(0) \) as a function of \( \tau \) on a log-log scale for \( r = 0.80, 0.85, 0.90, \) and \( 0.95 \). 

Next, we investigate the velocity autocorrelations for the initial condition. We plot \( \bar{C}(\tau_w, \tau) \) against \( \tau \) for \( \tau_w = 0 \) on a linear-log scale, considering the corresponding values of \( r = 0.80, 0.85, 0.90, \) and \( 0.95 \). From Figure 3, it can be observed that for a constant coefficient of restitution, the system with strong dissipation (\( r = 0.80 \)) exhibits a slow decay, indicating a stronger memory of previous velocities compared to the case of weak dissipation (\( r = 0.90 \)). In Figure 3 and 4, for \( r = 0.90 \) and \( r = 0.85 \), the system shows similar decay behavior across all values of dissipation.
In Figures 3 and 4, we examine \( \bar{C}(\tau_w, \tau) \) versus \( \tau \) for \( \tau_w > 0 \) (\( \tau_w = 10, 50, 100, \) and \( 200 \)) on a linear-log scale. In all figures, the decay of \( \bar{C}(\tau_w, \tau) \) becomes slower as the waiting times increase, demonstrating a dependence on \( \tau_w \). This behavior clearly indicates that our system exhibits aging properties, similar to that of the system with a constant coefficient of restitution. However, the decay rate is faster compared to the system with a constant coefficient of restitution. This difference arises because the system is not evolving with a constant coefficient of restitution; instead, the coefficient of restitution decreases as the system evolves due to the cooling of the granular gas. This results in a slower process of correlation among the velocities.
\section{\label{sec:summry} SUMMARY AND CONCLUSION}
This study investigates the aging behavior of the velocity autocorrelation function in a granular gas consisting of uniformly heated hard sphere particles, characterized by a coefficient of restitution \( r = 0.80, 0.85, 0.90, \) and \( 0.95 \). The results are compared with previously reported findings for systems with a constant coefficient of restitution across various values of \( r \).
For the case of \( \tau_w = 0 \), the velocity autocorrelation function \( \bar{C}(\tau_w, \tau) \) exhibits distinct behaviors for different restitution coefficients. However, when examining \( \tau_w > 0 \), the decay patterns of \( \bar{C}(\tau_w, \tau) \) for all restitution values become similar to those observed in systems with a constant \( r \). Notably, the rate of decay for the variable \( r \) is consistently faster than that for a constant \( r \). 
At early times, the normalized autocorrelation function decays exponentially, while in the later stages, the emergence of velocity correlations leads to a slower-than-exponential decay. This behavior indicates a significant dependence on the waiting time \( \tau_w \), which highlights the aging properties of the system. Furthermore, it is observed that at larger waiting times and under conditions of strong dissipation, the decay becomes notably slow. This suggests that the system retains a stronger memory of previous velocities compared to scenarios with weaker dissipation.
In conclusion, our findings provide valuable insights into the dynamic properties of granular gases, emphasizing the effects of the coefficient of restitution on aging behavior. The results suggest that systems with decreasing coefficients of restitution may exhibit complex dynamical behavior that differs significantly from those with constant coefficients. Future work could explore the implications of these findings on the macroscopic properties of granular materials and their practical applications in various engineering and industrial contexts.

\section*{\label{ack} ACKNOWLEDGEMENTS}
RFS acknowledges financial support from University Grants Commission in the form of Non-NET fellowships. He also wishes to acknowledge the computational facilities at the Department of Physics, JMI.

\section*{Declarations}
The authors declare that they have no known competing financial interests or personal relationships that could have appeared to influence the work reported in this paper.

\end{document}